\documentclass[aps,prd,showkeys,11pt,nofootinbib]{revtex4-2}

\usepackage{subcaption}
\usepackage{amsmath,amssymb,amsfonts,amsthm}
\usepackage{graphicx}
\usepackage{hyperref}

\def\barr{\begin{array}}
\def\earr{\end{array}}

\def\ben{\begin{equation}}
\def\een{\end{equation}}
\def\bs{\begin{subequations}}
\def\es{\end{subequations}}
\def\bena{\begin{eqnarray}}
\def\eena{\end{eqnarray}}
\def\im{{\rm i}}

\begin{document} 
\title{\bf Low-curvature quantum corrections from unitary evolution of de Sitter space}

\author{Steffen Gielen}
\affiliation{School of Mathematical and Physical Sciences, University of Sheffield, Hicks Building, Hounsfield Road, Sheffield S3 7RH, United Kingdom\footnote{Corresponding author}}
\email{s.c.gielen@sheffield.ac.uk}
\author{Rita B. Neves}
\affiliation{School of Mathematical and Physical Sciences, University of Sheffield, Hicks Building, Hounsfield Road, Sheffield S3 7RH, United Kingdom}
\email{rita.neves@sheffield.ac.uk}

\begin{abstract}
We study the quantum dynamics of de Sitter space formulated as a minisuperspace model with flat spatial hypersurfaces in unimodular gravity, both in the Wheeler--DeWitt approach and in loop quantum cosmology (LQC). Time evolution is defined naturally in unimodular time, which appears as conjugate to the cosmological (integration) constant. We show that requiring unitary time evolution ``resolves'' the de Sitter horizon where the flat slicing breaks down and leads to strong quantum effects there, even though locally nothing special happens at this surface. For a cosmological constant that is far below the Planck scale, loop quantum gravity corrections do not alter the main results in any substantial way. This model illustrates the fundamental clash between general covariance and unitarity in quantum gravity.
\end{abstract}

\keywords{Quantum cosmology, de Sitter space, unitarity, unimodular gravity}

\maketitle

\section{Introduction}

One of the main obstacles in our attempts to find a consistent quantum framework for gravity arises from the different concepts of time in general relativity and quantum mechanics. While general relativity allows general transformations from one coordinate frame to another, quantum mechanics relies on notions of time evolution and unitarity based on a more restricted concept of time. One immediate difficulty arises from the fact that the (bulk) Hamiltonian of a generally covariant theory vanishes due to diffeomorphism invariance, obstructing any interpretation of such a Hamiltonian as generator of time evolution. These issues have long been discussed in the literature as the {\em problem of time} \cite{Isham:1992ms,Kuchar:1991qf,Anderson:2012vk,Rovelli2023,Nanda:2023wne}. 

A popular strategy for addressing the lack of an absolute notion of time in quantum gravity is to choose a suitable degree of freedom as a ``relational'' clock for the remaining variables \cite{DeWitt:1962cg,Rovelli:1990ph}. Such a clock may not be valid globally if, e.g., it experiences a turnaround during its evolution, and one may want to restrict to the ideal case of a monotonically evolving variable \cite{PhysRevD.34.1040}. One then faces a {\em multiple-choice} problem \cite{Kuchar:1991qf} with no general expectation that the theories obtained for different clock choices would be equivalent (see, e.g., \cite{Malkiewicz:2016hjr} and the recent work \cite{Gielen:2020abd,Gielen:2021igw}).

There are many proposals for circumventing these conceptual and technical difficulties; in the following we will focus on using the clock degree of freedom introduced by passing from general relativity to unimodular gravity \cite{Unruh:1989db}. In unimodular gravity, the cosmological constant $\Lambda$ is an integration constant rather than a fundamental parameter of Nature. As such, this theory is classically equivalent to standard general relativity up to one additional global degree of freedom (see, e.g., \cite{Buchmuller:1988yn,Finkelstein:2000pg}). The conjugate variable to the ``total energy'' given by the value of $\Lambda$ then provides a global time variable, proportional to the total 4-volume in the past of a given spatial slice \cite{Henneaux:1989zc}. Of course, this is still a particular choice of clock \cite{Kuchar:1991xd} but it has the advantage of being defined directly in terms of spacetime geometry, independently of any other degrees of freedom (in contrast to models in which specific matter is coupled \cite{Brown:1994py}, for example).

In this context of (quantum) unimodular gravity, we will study the simplest type of geometries: homogeneous and isotropic Friedmann--Lema\^{i}tre--Robertson--Walker (FLRW) spacetimes. Classically, without matter and for a given value of the cosmological integration constant, the unique solution is given by either Minkowski, de Sitter, or anti-de Sitter space. Hence, the quantum theory is expected to be rather simple, only introducing the novelty of allowing for superpositions of the cosmological constant and hence superpositions of these distinct classical spacetimes. However, there is an important additional subtlety in that these spacetimes can be represented as different types of FLRW geometry using either closed, open or flat slices. Even though classically these apparently different cosmologies represent the same spacetime, crucial properties of the resulting quantum theories, even in this simplest possible case, can be sensitive to the choice of foliation, adding additional ambiguity beyond those related to a choice of time coordinate or clock.

In the example of de Sitter spacetime that we will focus on here, the crucial difference between the different slicings is that a closed slicing is global (it covers the entire spacetime) while the flat slicing only covers half of the spacetime; there is a horizon at which the foliation breaks down (see, e.g., \cite{Mukhanov}, for details on the different foliations of de Sitter spacetime). This horizon can be reached in finite time along past-directed geodesics, so that the spacetime becomes geodesically incomplete. In the Hamiltonian picture, one would say that past evolution terminates in finite time. Requiring that the quantum theory remains unitary means that the quantum evolution extends beyond this horizon, necessitating strong departures from the classical evolution in a regime where locally nothing special happens and no local observer would notice the presence of a horizon. Hence, it seems one has to give up either general covariance or unitarity (see also \cite{Gielen:2022tzi}).

The arguments presented here are somewhat reminiscent of discussions in the context of a possible breakdown of unitarity in black hole evaporation \cite{Hawking:1976ra}. For the black hole, requiring global unitarity, locality and general covariance -- nothing special should happen where curvature is small -- leads to a paradox  one possible resolution of which, the ``firewall'' idea \cite{Almheiri:2012rt}, likewise suggests that the black hole horizon becomes a region of strongly non-classical behaviour. As in many other contexts, there are important similarities between the black hole and de Sitter horizons (most famously, de Sitter horizons show thermodynamic behaviour \cite{Gibbons:1976ue}) but important differences too; while the black hole horizon has a unique ``teleological'' definition, the de Sitter horizon is observer (or foliation) dependent. Nevertheless, the simpler de Sitter case poses the same fundamental questions about the meaning of locality, unitarity and general covariance in quantum gravity. See also \cite{Gielen:2024lpm} for a recent study of a quantum planar black hole model using methods similar to those applied here, where quantum departures from the classical solution are seen both at the singularity and at the horizon of the black hole; the results for the horizon are similar to ours here.

Assuming FLRW symmetry means the geometric degrees of freedom are reduced to just the scale factor, here seen as evolving in unimodular time. The resulting quantum theories are simple and can be defined straightforwardly. We will see how unitary evolution enforces a ``resolution'' of the de Sitter horizon, replacing it by a highly quantum region that interpolates between two classical solutions each associated to one half of de Sitter spacetime. We will work both in a traditional Wheeler--DeWitt quantisation and in LQC, where many results show resolution of classical singularities through loop quantum gravity effects \cite{Ashtekar:2011ni,Banerjee:2011qu,Li:2023dwy}, so one could ask whether that approach also shows ``fake singularity resolution''. Specifically, we will follow the proposal of \cite{Chiou:2010ne} for defining LQC for unimodular gravity. We will find  agreement with the general conclusions of that article in that loop quantum corrections are small assuming that the cosmological constant is far away from the Planck scale, so the qualitiative features are similar to those of the Wheeler--DeWitt theory. The  fact that the quantum evolution continues through a highly quantum region replacing the classical horizon was however missed in \cite{Chiou:2010ne} due to some errors in the analysis, as we will clarify later.

In section \ref{sec:class} we summarise the classical formalism of unimodular gravity for FLRW geometries and written in Ashtekar--Barbero variables. Section \ref{sec:wdw} discusses the Wheeler--DeWitt quantisation of the model, a choice of Gaussian state and derivation of expectation values and standard deviations in the volume. We show our main result, namely strong quantum effects at the classical horizon. Section \ref{sec:lqc} extends the results to LQC, where we encounter some new features such as a maximum value for the cosmological constant and a quantum correction to $\Lambda$ coming from fundamental discreteness. Qualitative features of our solutions are similar to the Wheeler--DeWitt case. We conclude in section \ref{sec:concl}. Appendix \ref{app:calcs for generic epsilon} contains some technical details of the LQC calculations.

\section{Classical dynamics}
\label{sec:class}

In this section we briefly introduce the classical formalism of a ``parametrised'' version of unimodular gravity, reduced to FLRW spacetimes.  Our exposition is similar to the one of \cite{Chiou:2010ne}, and in particular based on the Ashtekar--Barbero variables used in loop quantum gravity and LQC. In the symmetry-reduced setting, these variables arise from a canonical transformation of the usual metric variables, with the additional novelty that they also include an orientation factor so that the volume variable (which we will mainly focus on in the following) can be positive or negative.

In unimodular gravity, fixing the determinant of the metric in the action leads to a theory of gravity in which only the trace-free part of the Einstein equations is imposed, and (after using the Bianchi identities) the cosmological constant emerges as an unspecified constant of integration in the missing trace equation (see, e.g., \cite{Ellis:2010uc} on the trace-free Einstein equations and \cite{Jirousek:2023gzr,Alvarez:2023juz} for recent reviews). The group under which the theory is invariant reduces to (volume-preserving or transverse) coordinate transformations that preserve the determinant of the metric. One can restore full diffeomorphism invariance by introducing additional fields, making the theory manifestly covariant, in a parametrised formalism for unimodular gravity \cite{Henneaux:1989zc}. In this setting, the action is given by
\begin{equation}
    S = \frac{1}{2\kappa} \int_{\mathcal{M}} {\rm d}^4x \left[ \sqrt{-g}(R-2\Lambda)+2\Lambda \partial_\mu T^\mu \right]\, 
\end{equation}
where $\kappa = 8\pi G$, and $\Lambda$ and $T^{\mu}$ are dynamical fields. Variations of the action with respect to $T^{\mu}$ imply that
\begin{equation}
    \partial_{\mu}\Lambda = 0
\end{equation}
and thus $\Lambda$ is identified with the cosmological constant. The field $T^{\mu}$ is a spacetime vector density that evolves according to the equation of motion obtained by varying the action with respect to $\Lambda$:
\begin{equation}\label{eq:unimodular condition}
    \sqrt{-g}=\partial_\mu T^\mu\, .
\end{equation}
For a spacetime manifold $\mathcal{M}=\Sigma\times\mathbb{R}$, where $\mathbb{R}$ denotes a time direction parametrised by a time coordinate $\tau$, the time evolution of $T^{\mu}$ is arbitrary except for the zero mode of $T^0$,
\begin{equation}\label{eq:Ttime}
    T(t_0) = \int_{\tau=t_0} {\rm d}^3 x\;T^0\,,
\end{equation}
which defines a global time coordinate related to the spacetime four-volume: $T(t_2)-T(t_1)$ is the total four-volume of the region bounded by the hypersurfaces $\tau=t_1$ and $\tau=t_2$. The remaining components of $T^\mu$ are pure gauge, needed to preserve locality while ensuring general covariance.

In the interest of later following the quantisation procedure of LQC, we adopt a Hamiltonian formalism and introduce connection-triad variables. We will summarise this procedure here, for more details see, e.g., \cite{Banerjee:2011qu,Ashtekar:2011ni}.  We start by performing the usual (3+1) decomposition writing the line element as
\begin{equation}
    {\rm d} s^2= (-N^2+N^iN_i)\, {\rm d}\tau^2 + 2N_i \, {\rm d}\tau\, {\rm d}x^i +q_{ij}\, {\rm d}x^i {\rm d}x^j
\end{equation}
where $N$ is the lapse function and $N^i$ the shift vector, and $q_{ij}$ is the three-metric induced on the spacelike slices $\Sigma$. Restricting to cosmological FLRW models, the line element is usually written in terms of a scale factor $a(\tau)$: in Cartesian coordinates,
\begin{equation}\label{eq:ds2 FLRW}
    {\rm d}s^2 = -N^2(\tau)\, {\rm d}\tau^2 + a^2(\tau)\,h_{ij} {\rm d}x^i {\rm d}x^j
\end{equation}
where
\begin{equation}
    h_{ij} = \delta_{ij} + k \frac{x_i x_j}{1-k\, x_k x^k}
\end{equation}
is a fiducial metric, and $k$ is spatial curvature, which takes the values $+1,0,-1$ for closed, flat or open slices, respectively. In the following we will focus on the flat case $k=0$.

LQC is an application of loop quantum gravity techniques to cosmological models. Instead of describing the phase space of general relativity with standard geometrodynamical variables (the spatial metric $q_{ij}$ and extrinsic curvature $K_{ij}$), one uses a reformulation of gravity in terms of an $SU(2)$ connection (known as Ashtekar--Barbero connection) $A^a_i$ and its canonically conjugate densitised triad $E^i_a$ (see, e.g., \cite{Rovelli:2004tv}),
\begin{equation}
A^a_i = \Gamma^a_i[E] + \gamma K^a_i\,, \quad E^i_a = \frac{1}{2}\epsilon^{ijk}\epsilon_{abc} e^b_j e^c_k
\end{equation}
where $\Gamma^a_i$ is the torsion-free Levi-Civita connection associated to $E$, $K^a_i$ is the extrinsic curvature 1-form, $e^b_j$ is a spatial triad such that $q_{ij}=\delta_{ab}\,e^a_i e^b_j$, and $\gamma$ is a free parameter known as the Barbero--Immirzi parameter. In the case of flat FLRW cosmology, these fields take the form 
\begin{equation}
A^a_i = V_0^{-1/3}c(\tau)\, {}^o e^a_i\,,\qquad E_a^i = V_0^{-2/3}\sqrt{h}\,p(\tau)\,{}^o e_a^i
\label{AEdef}
\end{equation}
where ${}^o e_i^a$ is a fiducial (co-)triad adapted to the metric $h_{ij}$, i.e., $h_{ij}=\delta_{ab}({}^o e^a_i)({}^o  e^b_j)$, and ${}^o e_a^i$ is the corresponding inverse triad. Notice that compared to the metric formulation there is now a new symmetry under the orientation reversal of the fiducial triad. The quantity $V_0$ corresponds to the coordinate volume $V_0=\int {\rm d}^3 x\;\sqrt{h}$ of a finite ``fiducial cell'' one wants to study. This cell could be the entire spatial slice if one assumes compact topology, such as a 3-torus, or a compact subset in case of infinite spatial slices. In the case of non-compact topology, the introduction of a fiducial cell is required to obtain regular expressions since otherwise the total action or Hamiltonian would diverge. The powers of $V_0$ appearing in (\ref{AEdef}) are needed to ensure that none of the properties of the resulting Hamiltonian system depend on the value of $V_0$, which is coordinate dependent. More details on these technical aspects and the general construction can be found in \cite{Banerjee:2011qu,Ashtekar:2011ni}.

The variables $c$ and $p$ form a canonical pair $\lbrace c,p \rbrace = \kappa \gamma/3$. For simplification of future calculations, it is then conventional to switch to a new canonical pair (we drop the explicit dependence on $\tau$ from now on)
\begin{equation}\label{bnudef}
    b \equiv \frac{c}{\sqrt{|p|}}\, , \qquad \nu \equiv \textrm{sign}(p) \frac{4|p|^{3/2}}{\kappa \hbar \gamma}\,, \qquad 
    \lbrace b,\nu \rbrace = \frac{2}{\hbar}\,.
\end{equation}
These are related to the more familiar variables of (\ref{eq:ds2 FLRW}) by
\begin{equation}
 |\nu| = \frac{4}{\kappa\hbar\gamma} a^3 V_0\,,\qquad b = \pm \gamma\frac{\dot{a}}{aN}\,.
 \label{nubexplain}
\end{equation}
Symmetry under the reversal of triad orientation   ${}^o e^a_i\rightarrow -{}^o e^a_i$ is now translated to the symmetry under the change of signs of $\nu$ and $b$. Notice that $|\nu|$ is proportional to the physical three-volume while $b$ is proportional to the Hubble rate $\dot{a}/(aN)$; since the proportionality factors involve $\gamma$, the combinations $\gamma|\nu|$ and $b/\gamma$ are directly related to the physically relevant variables in cosmology. Again, we stress that the overall sign of $\nu$ is interpreted as related to orientation of the triad.

In the extension of the formalism to the unimodular setting, the fields $T^\mu$ and $\Lambda$ remain unchanged. FLRW symmetry implies that all $T^{\mu}$ vanish except for $T^0=T^0(\tau)$.  Then, the global time (\ref{eq:Ttime}) is simply given by $T=T^0V_0$, and forms a canonical pair with the cosmological constant:
\begin{equation}
    \lbrace T, \Lambda \rbrace = \kappa\, .
\end{equation}

The final consequence of homogeneity and isotropy is that spatial diffeomorphisms are trivially satisfied, and the Hamiltonian reduces to the product of the lapse $N$ with a function $C$ given by
\begin{equation}
    H= N\left(-\frac{3\hbar}{4\gamma} b^2 |\nu|+\frac{\gamma \hbar}{4}\Lambda|\nu|\right) \equiv N\, C\,.
\end{equation}
The function $C$ is constrained to vanish through the equation $\partial H/\partial N = 0$, and it is a generator of time reparametrisations on the constraint hypersurface in phase space. In the interest of choosing $T$ as the internal clock, in the following we will fix the lapse to $N = \frac{4}{\kappa\gamma \hbar |\nu|}$. Thus, we obtain the classical Hamiltonian
\begin{equation}\label{eq:classicalC}
    H = -\frac{3}{\kappa\gamma^2}b^2 + \frac{\Lambda}{\kappa} \approx 0
\end{equation}
which is again constrained to vanish. (With this lapse choice, indeed $\dot{T}=\{T,H\}=1$.)

Implementing (\ref{eq:classicalC}) at the quantum level is the basis of quantum cosmology both for Wheeler--DeWitt and LQC-type quantisations. However, in the quantisation of LQC, there is no operator that directly represents the connection and so no $b$ operator. Instead, fundamental variables must be represented as finite holonomies of the connection and fluxes of the densitised triad. Following the improved dynamics prescription of LQC \cite{Ashtekar:2006wn}, the holonomies of the connection are taken along straight lines in the fundamental representation of $SU(2)$ with a length such that they enclose a physical area $\lambda^2$, where $\lambda^2$ is chosen to represent the minimum non-vanishing eigenvalue of the loop quantum gravity area operator. In the variables ($\nu,b$), the holonomies produce a constant shift in $\nu$, rather than acting as a derivative in $\nu$ as $b$ itself would. We will present details of this  theory shortly and compare it with the more conventional Wheeler--DeWitt quantisation.

We can obtain classical trajectories by computing Poisson brackets of variables with the Hamiltonian. $\Lambda$ and $b$ are constants of motion related through \eqref{eq:classicalC}. We are then particularly interested in tracking the physical volume, which is proportional to $|\nu|$: setting an integration constant corresponding to a shift in $T$ to zero, the classical solution is
\begin{equation}\label{eq:|nu|(T) classical}
    \left|\nu(T)\right| = \frac{4}{\hbar\kappa\gamma}\sqrt{3\Lambda} |T|\,.
\end{equation}
The classical trajectories are therefore those of a contracting universe for $T<0$ and expanding universe for $T>0$. We also note that with this choice of clock the volume vanishes at a finite time, $T = 0$. This is simply a coordinate singularity since the curvature is constant, and the point $T=0$ is of no special relevance to the dynamics. If we had chosen cosmic time $t$ instead ($N=1$), we would have found that $\left|\nu(T)\right| \sim e^{\sqrt{3\Lambda}t}$, and vanishing volume is only reached in the infinite past. This seemingly harmless coordinate singularity will become crucial when we demand unitarity in time $T$.

\section{Wheeler--DeWitt quantisation}
\label{sec:wdw}

Let us consider first the more straightforward Wheeler--DeWitt quantisation of the theory defined by the Hamiltonian constraint (\ref{eq:classicalC}). We take the representation where $\hat{T}$ and $\hat{\nu}$ act by multiplication and their conjugates by derivative: $\hat{\Lambda}=-\im\hbar\kappa\frac{\partial}{\partial T}$, $\hat{b}=2\im \frac{\partial}{\partial\nu}$ (the unusual factors for $\hat{b}$ arise from the conventions set in (\ref{bnudef})). 

The action of the quantum representation of the constraint on states $\Psi(\nu,T)$ leads to the Wheeler--DeWitt equation which here is of Schr\"odinger form
\begin{equation}
    \im\hbar \frac{\partial\Psi(\nu,T)}{\partial T} = \frac{12}{\kappa \gamma^2} \frac{\partial^2}{\partial \nu^2}\Psi(\nu,T).
\end{equation}
This is simply a free particle in one dimension, with general solution
\begin{equation}
\Psi(\nu,T) =\int_0^\infty  {\rm d}\Lambda\;e^{\im\frac{\Lambda T}{\hbar\kappa}}\left(A(\Lambda) e^{\im\sqrt{\frac{\Lambda \gamma^2}{12}}\nu} + B(\Lambda) e^{-\im\sqrt{\frac{\Lambda \gamma^2}{12}}\nu}\right)\,.
\end{equation}
While the solution depends on the free parameter $\gamma$, it is only in the combination $\gamma\nu$ which is related to the physical three-volume (see (\ref{nubexplain})).  Also recall that the sign of $\nu$ corresponds to a triad orientation and geometric observables are insensitive to this sign. The quantum Hamiltonian constraint commutes with the parity operator $\hat\Pi$ sending $\Psi(\nu,T)$ to $\Psi(-\nu,T)$. It is then conventional to assume that the wavefunction is even under parity \cite{Ashtekar:2011ni,Chiou:2010ne}, and we will make the same assumption here.

With this symmetry assumption, our general solution simplifies to
\begin{equation}
    \Psi(\nu,T) = \int_{0}^{\infty} {\rm d}\Lambda\; A(\Lambda)\; e^{i\frac{\Lambda T}{\hbar\kappa}}\cos\left(\sqrt{\frac{\Lambda \gamma^2}{12}}\nu\right)\,.
\end{equation}
We assume the natural Schr\"odinger $L^2$ inner product
\begin{equation}\label{eq:expWdW}
    \left( \Psi_1, \Psi_2 \right) = \int_{-\infty}^{\infty} {\rm d}\nu\; \bar{\Psi}_1(\nu,T) \Psi_2(\nu,T)
\end{equation}
which evaluates to
\begin{equation}\label{eq:normWdW}
    \left( \Psi, \Psi \right) = \frac{4\pi}{\gamma} \int_{0}^{\infty} {\rm d}\Lambda\;|A(\Lambda)|^2 \sqrt{3 \Lambda}\ .
\end{equation}

Since the classical volume of spatial slices is proportional to $|\nu|$, we are particularly interested in tracking the expectation value of $\widehat{|\nu|}$,
\begin{equation}\label{eq:expv}
    \left( \Psi, \widehat{|\nu|}\Psi \right) = \int_{-\infty}^{\infty} {\rm d}\nu\; |\nu| \bar{\Psi}(\nu,T) \Psi(\nu,T)\,.
\end{equation}
An attempt to evaluate this expectation value in terms of the amplitude $A(\Lambda)$ requires the knowledge of the Fourier transform of $|x|$, which is usually (e.g., \cite{1130282270434563200}) given as
\begin{equation}\label{eq:Fourier?}
 \int {\rm d}x \;|x|\,e^{\im k x}=-\frac{2}{k^2}
\end{equation}
and so formally we could write 
\begin{equation}\label{eq:formalabsnu}
\left( \Psi, \widehat{|\nu|}\Psi \right) = -\frac{12}{\gamma}\int_{0}^{\infty} {\rm d}\Lambda\int_{0}^{\infty} {\rm d}\Lambda'\;\overline{A(\Lambda)}A(\Lambda')\left(\frac{1}{\left(\sqrt{\Lambda}+\sqrt{\Lambda'}\right)^2}+\frac{1}{\left(\sqrt{\Lambda}-\sqrt{\Lambda'}\right)^2}\right)\,.
\end{equation}
The resulting integral clearly has singularities for $\Lambda=\Lambda'$ which require regularisation.\footnote{In the Fourier transform (\ref{eq:Fourier?}), the expression $1/k^2$ should really be understood as a homogeneous distribution defined, e.g., as acting on test functions via \cite{gelfand} $\underline{k}^{-2}(\phi) = \int {\rm d}k\,\frac{\phi(k)-\phi(0)-k \phi'(0)}{k^2}$. This definition of the Fourier transform could be used to make (\ref{eq:formalabsnu}) well-defined. We thank Sofie Ried for clarifying this to us.} 

As a concrete example, we focus on simple states defined by a Gaussian profile in $\sqrt{\Lambda}$,
\begin{equation}\label{eq:GaussianLambda}
    A(\Lambda) = N e^{-\frac{(\sqrt{\Lambda}-\sqrt{\Lambda_0})^2}{2 \sigma^2}}
\end{equation}
where $N$ is a normalisation constant. Then we can perform a change of integration variable to $u = \sqrt{\Lambda}$ and write the states as
\begin{equation}\label{eq:WdWstates int u}
    \Psi(\nu,T) =  N \int_{0}^{\infty} {\rm d}u\, u\, \left(e^{-f(T) u^2 + g(\nu) u -\frac{\Lambda_0}{2\sigma^2}} + e^{-f(T) u^2 + \bar{g}(\nu) u -\frac{\Lambda_0}{2\sigma^2}}\right),
\end{equation}
where the bar denotes complex conjugation and we have defined
\begin{align}
    f(T) &\equiv \frac{1}{2\sigma^2}-\im \frac{T}{\hbar \kappa}\, ,\label{eq:f(T)}\\
    g(\nu) &\equiv \frac{\sqrt{\Lambda_0}}{\sigma^2} + \im \sqrt{\frac{\gamma^2}{12}}\nu\, .
\end{align}
The $u$ integral can be performed exactly, yielding (with suppressed arguments in $f$ and $g$)
\begin{equation}
    \Psi(\nu,T) = N \frac{e^{-\frac{\Lambda_0}{2\sigma^2}}}{2f}\left\{4 + \sqrt{\frac{\pi}{f}}\left(g e^{\frac{g^2}{4f}}\left[1+\text{erf}\left(\frac{g}{2\sqrt{f}}\right)\right]+\bar{g}e^{\frac{\bar{g}^2}{4f}}\left[1+\text{erf}\left(\frac{\bar{g}}{2\sqrt{f}}\right)\right]\right)\right\}
    \label{eq:3.13}
\end{equation}
where $\text{erf}(x)=\frac{2}{\sqrt{\pi}}\int_0^x {\rm d}y\;e^{-y^2}$ denotes the error function.

It may be useful to consider that due to the peaked nature of the profile, the integral \eqref{eq:WdWstates int u} can be well approximated by taking it over the whole real line, resulting in
\begin{equation}
    \Psi(\nu,T) \approx N \frac{e^{-\frac{\Lambda_0}{2\sigma^2}}}{2 f(T)}\sqrt{\frac{\pi}{f(T)}} \left(g(\nu)\ e^{\frac{g^2(\nu)}{4f(T)}}+\bar{g}(\nu)\ e^{\frac{\bar{g}^2(\nu)}{4f(T)}} \right)\,.
    \label{approximatePsi}
\end{equation}

\begin{figure}
    \centering
    \includegraphics[width=0.9\textwidth]{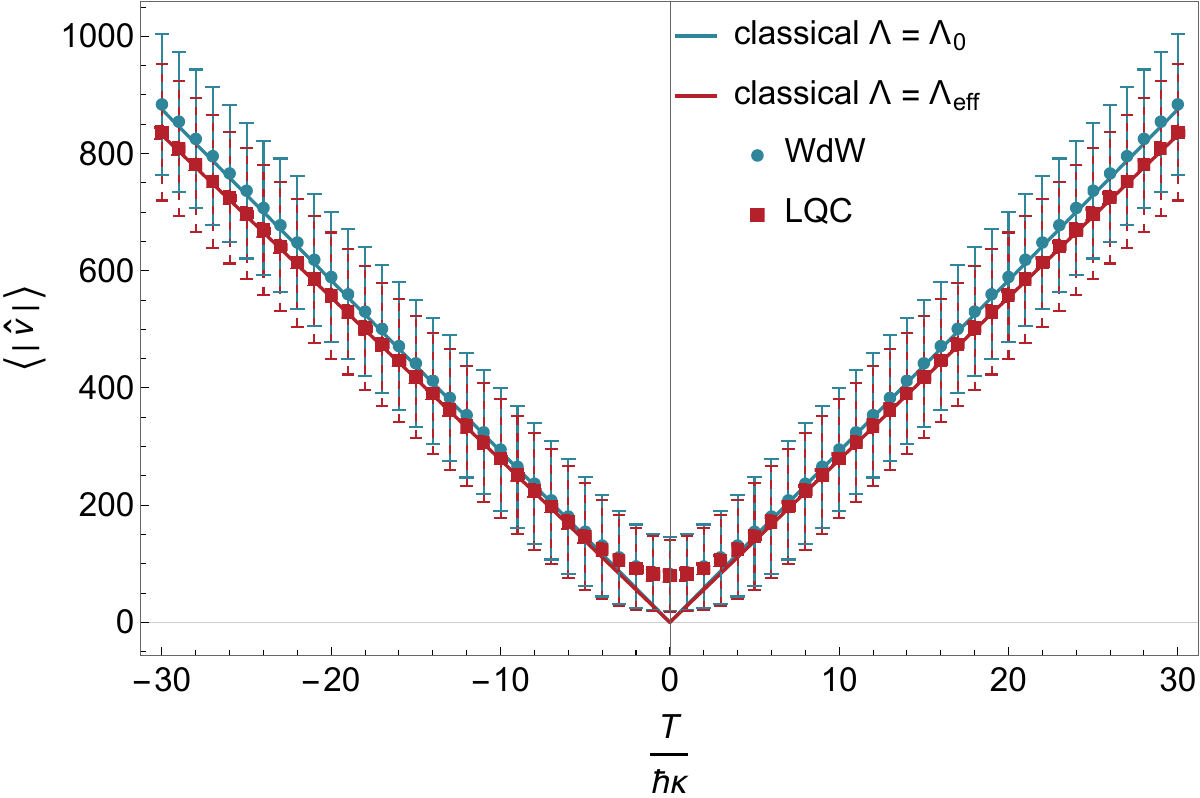}
    \caption{Expectation value of $|\hat{\nu}|$ as a function of $T$ for Wheeler--DeWitt theory and LQC for a state characterised by a Gaussian profile \eqref{eq:GaussianLambda} with $\Lambda_0=1$ and $\sigma = 0.1$, shown with the standard deviation and corresponding classical trajectories. Here $\hbar=G=1$ so that $\kappa=8\pi$; we also set $\lambda = \sqrt{4\sqrt{3}\pi \gamma}$ and $\gamma = 0.2375$ as discussed below (\ref{modifiednu}). Then $\Lambda_p\approx 10.29$.}
    \label{fig:expv}
\end{figure}

One can then perform the integral \eqref{eq:expv} numerically. Choosing some example values for the parameters of the Gaussian, we compute this expectation value and plot it alongside the classical trajectories in figure \ref{fig:expv}. We also include error bars showing the standard deviation in $|\nu|$, $\Delta|\nu|=\sqrt{\langle\widehat{|\nu|^2}\rangle - \langle\widehat{|\nu|}\rangle^2}$. As expected and alluded to in the introduction, the quantum trajectory deviates from the classical one close to the de Sitter horizon, where it interpolates between the two classical solutions corresponding to different halves of the whole de Sitter spacetime. This is a consequence of imposing unitarity in a region where the classical solution goes to zero.

This situation is analogous to computing the expectation value of the absolute value of the position of a wavepacket for a free particle moving in one direction. The position variable $x$ can take on positive and negative values, as can $\nu$ in our case. For a wavepacket, $\langle\hat{x}\rangle$ would follow the classical trajectory and eventually reach and cross $x = 0$. However, once we consider $|\hat{x}|$, the negative and positive $x$ parts of the wavefunction both contribute positively to $\langle|\hat{x}|\rangle$, and $\langle |\hat{x}| \rangle$ can never reach exactly 0. The trajectory of $\langle |\hat{x}| \rangle$ therefore closely tracks the classical trajectory of $|x|$ when far from $x=0$, but as it approaches this point (of no special significance in the classical trajectory), it necessarily deviates from it. Unitarity does not allow for the quantum trajectory to simply end there and $\langle |\hat{x}| \rangle$ ``bounces'' close to $x=0$, resulting in the effect we observe in figure \ref{fig:expv}. Here, because of the symmetry $\Psi(\nu,T)=\Psi(-\nu,T)$, our state contains two wavepackets corresponding to an expanding and a contracting de Sitter solution; this is the equivalent of a superposition of two wavepackets that are mirror images of each other, moving in opposite directions. Both of them will contribute to $\langle |\hat{x}| \rangle$ equally.

While this is all clear from the analogy with a free particle and not a surprise from the quantum point of view, the suggested implications for quantum gravity seem dramatic. The result suggests that the classical horizon (which is foliation-dependent with nothing special happening locally) is replaced by a quantum region that deviates strongly from the classical solution: the ratio of quantum and classical volumes $\langle|\hat\nu(T)|\rangle/|\nu_{{\rm cl}}(T)|$, where $\nu_{{\rm cl}}(T)$ is the classical solution corresponding to our chosen parameter values, evidently diverges as $T\rightarrow 0$. This conclusion seems inevitable for any choice of semiclassical state, regardless of how sharply peaked it was chosen to be away from $T=0$; the qualitative features seen in figure \ref{fig:expv} are not sensitive to the details in the state, as long as it is semiclassical. Away from $T=0$, expectation values closely follow the classical solution, while there are strong departures near $T=0$. 

\begin{figure}
    \centering
    \includegraphics[width=0.82\textwidth]{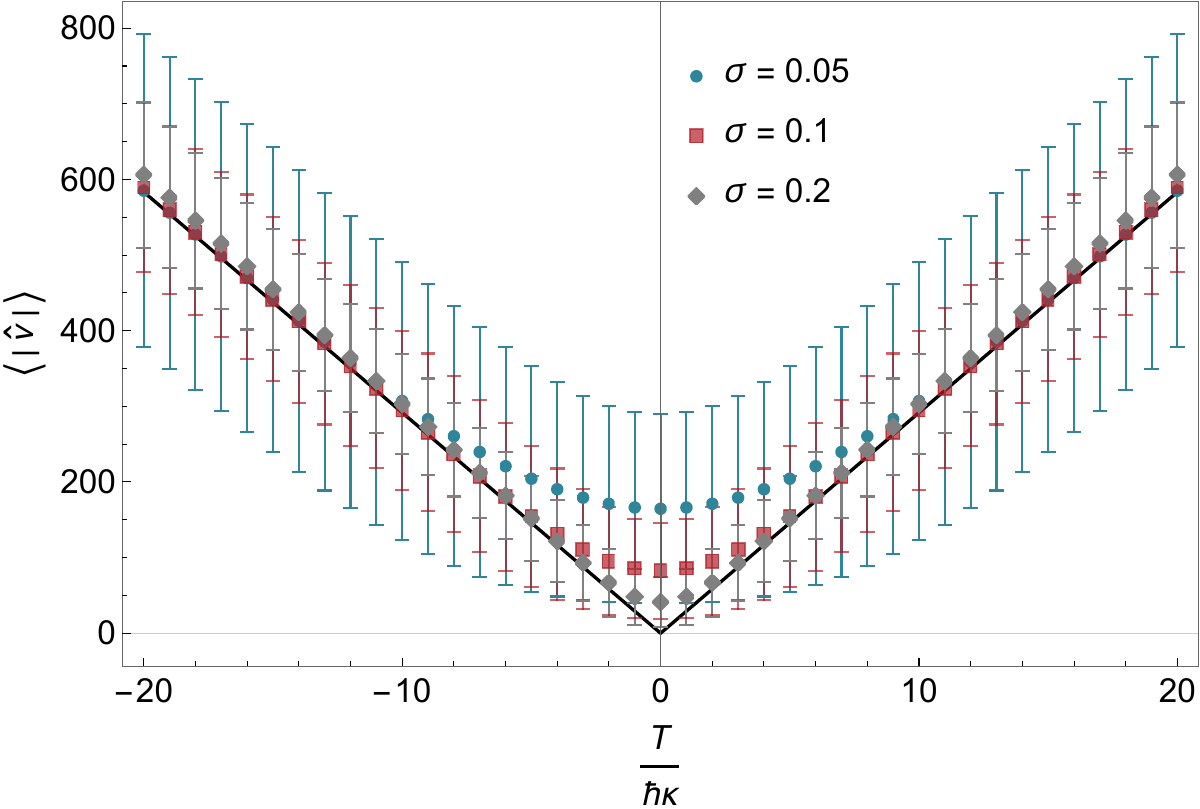}
    \caption{Expectation value of $|\hat{\nu}|$ as a function of $T$ for Wheeler--DeWitt theory for a state with Gaussian profile \eqref{eq:GaussianLambda} with different values for $\sigma$ and $\Lambda_0=1$, plotted along with standard deviations and the corresponding classical trajectory with $\Lambda=\Lambda_0$ (black lines).}
    \label{fig:sigmacomp}
\end{figure}

More quantitatively, we can compute $\langle|\hat\nu(0)|\rangle$ using the approximation (\ref{approximatePsi}). For $\sigma\ll\sqrt{\Lambda_0}$,
\begin{equation}
\langle|\hat\nu(0)|\rangle \approx \sqrt{\frac{3}{\pi}}\, \frac{2}{\gamma\sigma}\,.
\label{nuatzero}
\end{equation}
The scaling of the minimal value of the volume in the quantum theory with $1/\sigma$ agrees with that seen in similar constructions in quantum cosmology such as \cite{GrybThebault}. In particular, the precise value depends on the details of the state. However, given that the classical volume goes to zero here, we would see any non-zero value as a strong quantum departure from the classical solution. The dependence on $\sigma$ is more explicitly illustrated in figure \ref{fig:sigmacomp}. We have checked in particular that the results for $\langle|\hat\nu(0)|\rangle$ match with the analytical approximation (\ref{nuatzero}).

\begin{figure}
    \centering
    \includegraphics[width=0.7\textwidth]{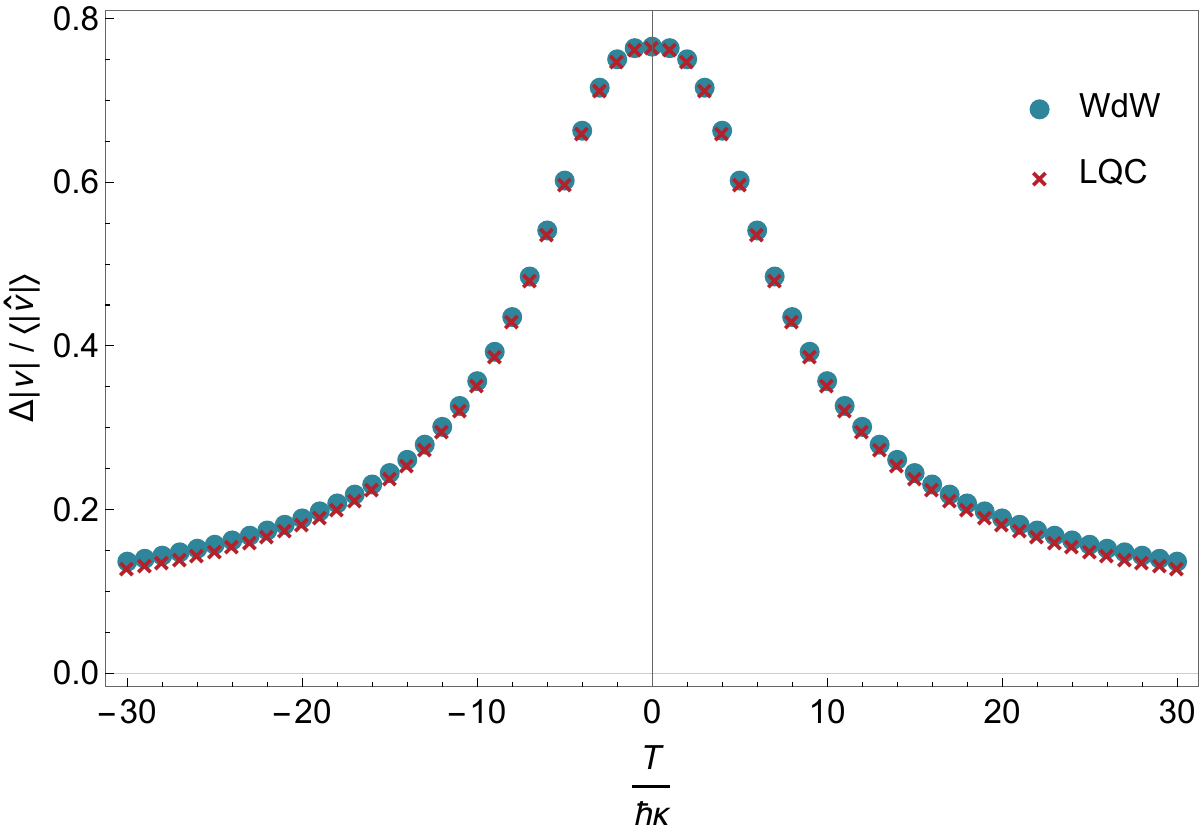}
    \caption{Relative standard deviation $\Delta|\nu|/\langle|\hat\nu|\rangle$ as a function of $T$ for Wheeler--DeWitt theory and LQC, with state parameters chosen as in figure \ref{fig:expv}.}
    \label{fig:relvariance}
\end{figure}

As a last point, we notice that the dispersion in $|\nu|$ grows away from $T=0$, as indicated by an increasing length of the error bars shown in the previously discussed figures. However, if one looks at the relative standard deviation $\Delta|\nu|/\langle|\hat\nu|\rangle$, a dimensionless ratio which might be more meaningful to study, one can see that it is actually greatest around the quantum bounce and then falls off as the volume grows towards the future or past. This is also illustrated in figure \ref{fig:relvariance}. 

In contrast to the results presented here, the analysis of \cite[Sec.~VI]{Chiou:2010ne} suggests that the expectation value $\langle|\hat\nu|\rangle$, and hence the volume of spatial slices, always remains linear in $T$ following the classical solution (\ref{eq:|nu|(T) classical}) for positive $T$, so it does eventually go to zero (and then turns negative, which is impossible). Closer inspection of that paper (e.g., Eq.~(6.7)) shows that the authors actually calculate the expectation value of $\hat\nu$ rather than of $|\hat\nu|$, which would be zero at all times due to the assumed symmetry.

\section{Loop quantum cosmology quantisation}
\label{sec:lqc}

In this section we investigate the impact of moving from a standard Wheeler--DeWitt quantisation to an LQC model, very similar to what was proposed in \cite{Chiou:2010ne}. LQC mimics loop quantum gravity effects in effectively replacing the continuum of space by a lattice; the connection $b$ cannot directly be represented as an operator $\hat{b}$ but must be approximated by finite holonomies $\widehat{e^{\im \lambda b/2}}$ where the limit $\lambda\rightarrow 0$ is not well-defined in the theory. We refer the interested reader to the reviews \cite{Ashtekar:2011ni,Banerjee:2011qu} for more on the motivations and technical aspects of such an approach. Here we are mainly interested in the implications for the quantum resolution of the de Sitter horizon that we have observed, given that resolution of classical singularities (usually the curvature singularity at the Big Bang) is perhaps the main result of LQC.

As replacing the connection by holonomies is similar to putting the theory on a lattice, it introduces discretisation ambiguities. There are additional ambiguities, e.g.,  ``inverse triad'' corrections coming from the fact that inverse powers of the volume variable $|\nu|$ must also be regularised in loop quantum gravity. As a result, using different prescriptions leads to different, inequivalent models (see, e.g., \cite{Yang:2009fp,Assanioussi:2018hee}). Qualitative features of these models, such as singularity resolution by a bounce, are more generic and can already be obtained in simplified solvable models \cite{Ashtekar:2007em}, which we will also restrict ourselves to. Again this is very similar to the treatment of \cite{Chiou:2010ne}.

In these simplified models, the main change compared to a standard quantisation is to make the replacement
\begin{equation}
    b \rightarrow \frac{\widehat{\sin(\lambda b)}}{\lambda}\, ,
\end{equation}
where $\lambda$ can in general be a constant or a function of other variables, chosen through physical requirements and a heuristic connection with loop quantum gravity. The choice of constant $\lambda$ is usually preferred \cite{Ashtekar:2011ni}. 

Let us first point out that such a replacement would modify the dynamics even in the classical limit. Indeed, if we replace the constraint (\ref{eq:classicalC}) by a loop-modified version
\begin{equation}
    H = -\frac{3}{\kappa\gamma^2}\frac{\sin^2(\lambda b)}{\lambda^2} + \frac{\Lambda}{\kappa} \approx 0\,,
\end{equation}
the classical solution (\ref{eq:|nu|(T) classical}) is replaced by
\begin{equation}
    \label{modifiednu}
    \left|\nu(T)\right| = \frac{4}{\hbar\kappa\gamma}\sqrt{3\Lambda-\lambda^2\gamma^2\Lambda^2}\, |T|\,.
\end{equation}
The $\lambda$ correction has the effect of ``screening'' the cosmological constant, so that the classical solution is now the one corresponding to $\Lambda_{{\rm eff}}=\Lambda-\frac{1}{3}\lambda^2\gamma^2\Lambda^2<\Lambda$. This is very similar to the effective dynamics obtained in a classical limit in \cite{Chiou:2010ne}. The correction depends on the product $\lambda\gamma$ which defines a new scale corresponding to discreteness of the geometry. Again, this can be seen as a free parameter but  is often fixed by identifying $\lambda^2$ with the minimum non-vanishing eigenvalue of the area operator in loop quantum gravity \cite{Rovelli:2004tv}, $\lambda^2 = 4 \sqrt{3}\pi G \gamma$, and taking $\gamma = 0.2375$ from a calculation of black hole entropy in loop quantum gravity \cite{Meissner:2004ju}. Requiring that $\sin^2(\lambda b) \leq 1$ implies that there are no solutions for $\Lambda>\Lambda_p\equiv \frac{3}{\gamma^2 \lambda^2}$, and so the LQC corrections introduce an upper limit for a positive cosmological constant allowed in the theory. If we associate the discretisation scale $\gamma\lambda$ to the Planck scale, $\Lambda_p$ denotes a Planckian cosmological constant. We then also have $\Lambda_{{\rm eff}}=\Lambda(1-\frac{\Lambda}{\Lambda_p})$.

In the representation where $\hat{\nu}$ acts by multiplication, the holonomy operator $\widehat{e^{\im \lambda b/2}}$ then produces a constant shift $-\lambda$ in $\nu$, and 
\begin{equation}
    \frac{\widehat{\sin(\lambda b)}}{\lambda}\Psi(\nu) = \frac{1}{2 \im\lambda}(\Psi(\nu-2\lambda)-\Psi(\nu+2\lambda))\,.
\end{equation}
Evidently, a formal continuum limit $\lambda\rightarrow 0$ reproduces the standard Wheeler--DeWitt quantisation we discussed previously.

The constraint (\ref{eq:classicalC}) is now represented as a difference equation
\begin{equation}
\label{eq:differenceWDW}
    \im \hbar \frac{\partial \Psi(\nu,T)}{\partial T} = \frac{3}{4 \kappa \gamma^2 \lambda^2}\left[\Psi(\nu-4\lambda,T)-2 \Psi(\nu,T) + \Psi(\nu+4\lambda,T)\right]\,.
\end{equation}
Because the wavefunction is only propagated in discrete steps in $\nu$, $\Psi(\nu,T)$ can be restricted to have support on $\mathcal{L}_\epsilon\times\mathbb{R}$, where $\mathcal{L}_\epsilon=\lbrace\epsilon \pm 4n\lambda,\, n \in \mathbb{Z}\rbrace$, $\epsilon \in [0,4)$. The choice $\epsilon=0$ is often the most interesting one since it would include the classical singularity $\nu=0$.

The physical inner product in this representation is then simply
\begin{equation}\label{eq:inner product LQC}
    \left( \Psi_1, \Psi_2 \right) = 4\lambda \sum_{\nu \in \mathcal{L}_\epsilon} \bar{\Psi}_1(\nu, T_0) \Psi_2(\nu,T_0).
\end{equation}
Again, we can see that the continuum limit is given by \eqref{eq:expWdW}.

 Taking the same separation ansatz as in the previous section, $\Psi(\nu,T) \sim e^{\im\frac{\Lambda T}{\hbar\kappa}}\Psi_{\Lambda}(\nu)$, yields
\begin{equation}
    2\left( 1 - 2\frac{\Lambda}{\Lambda_p} \right) \Psi_{\Lambda}(\nu) = \Psi_{\Lambda}(\nu-4\lambda) + \Psi_{\Lambda}(\nu+4\lambda)\,.
\end{equation}

This difference equation can be solved exactly with the general solution given by
\begin{equation}
    \Psi_{\Lambda}(\nu) = \left(1 - 2\frac{\Lambda}{\Lambda_p}  -2 \sqrt{\frac{\Lambda}{\Lambda_p} \left(\frac{\Lambda}{\Lambda_p}-1\right)} \right)^{\frac{\nu}{4\lambda}} c_1(\Lambda) + \left( 1 - 2\frac{\Lambda}{\Lambda_p}  + 2\sqrt{\frac{\Lambda}{\Lambda_p} \left(\frac{\Lambda}{\Lambda_p}-1\right)} \right)^{\frac{\nu}{4\lambda}} c_2(\Lambda).
\end{equation}
Demanding again $\Psi_{\Lambda}(-\nu) = \Psi_{\Lambda}(\nu)$ (invariance under triad reversal) leads to $c_1(\Lambda) = c_2(\Lambda)$. A normalised symmetric solution can then be rewritten as\footnote{In the limiting case $\Lambda=\Lambda_p$ we have $\alpha_\Lambda=\pi$ and $\beta_\Lambda=0$, so $\Psi_\Lambda(\nu)=(-1)^{\frac{\nu}{4\lambda}}$ from both expressions.}
\begin{equation}
\Psi_\Lambda(\nu)=\begin{cases} \cos\left(\frac{\nu}{4\lambda}\alpha_\Lambda\right)\,,\quad \Lambda \le \Lambda_p\,,
\\ (-1)^{\frac{\nu}{4\lambda}}\cosh\left(\frac{\nu}{4\lambda}\beta_\Lambda\right)\,,\quad \Lambda > \Lambda_p
\end{cases}
\end{equation}
with
\begin{equation}\label{eq:alphabetadef}
    \alpha_\Lambda \equiv \arctan\left(\frac{2\sqrt{\frac{\Lambda}{\Lambda_p}(1-\frac{\Lambda}{\Lambda_p})}}{1-2\frac{\Lambda}{\Lambda_p}}\right)\,,\quad \beta_\Lambda \equiv \log\left(2\frac{\Lambda}{\Lambda_p}- 1 - 2\sqrt{\frac{\Lambda}{\Lambda_p} \left(\frac{\Lambda}{\Lambda_p}-1\right)}\right)\,,
\end{equation}
where we define $\arctan(x)$ in the branch $[0,\pi]$ instead of the principal branch $[-\pi/2,\pi/2]$.\footnote{Note that $\alpha_\Lambda$ is simply the complex phase of $z = 1-2\frac{\Lambda}{\Lambda_p} + 2\im\, \sqrt{\frac{\Lambda}{\Lambda_p}(1-\frac{\Lambda}{\Lambda_p})}$ for $\Lambda \leq \Lambda_p$. Considering the signs of the real and imaginary parts of $z$, in the principal branch of $\arctan(x)$, one finds $\alpha_\Lambda = \arctan\left[\textrm{Im}(z)/\textrm{Re}(z)\right]$ for $\Lambda \leq \Lambda_p/2$, and $\alpha_\Lambda = \arctan\left[\textrm{Im}(z)/\textrm{Re}(z)\right]+ \pi$ for $\Lambda > \Lambda_p/2$.} The general symmetric solution to (\ref{eq:differenceWDW}) is 
\begin{equation}
    \Psi(\nu,T) = \int_0^{\infty} {\rm d}\Lambda\; A(\Lambda)\, e^{\im\frac{\Lambda T}{\hbar\kappa}} \,\Psi_\Lambda(\nu)\,.
\end{equation}

We have written the integration over the whole positive real line as in principle $\Lambda$ is allowed to take any positive value. However, we can show that only the solutions with $\Lambda\le \Lambda_p$ are normalisable with respect to the inner product \eqref{eq:inner product LQC} . Let us write the norm of a generic state as 
\begin{equation}\label{eq:norm}
    \left( \Psi, \Psi \right) = 4\lambda\int_0^{\infty} \int_0^{\infty} {\rm d}\Lambda\, {\rm d}\Lambda^{\prime}\, \overline{A(\Lambda)} A(\Lambda^{\prime})\, e^{\im\frac{\Lambda^{\prime}-\Lambda}{\hbar\kappa}T} \sum_{\nu \in \mathcal{L}_\epsilon} \bar{\Psi}_{\Lambda}(\nu) \Psi_{\Lambda^{\prime}}(\nu).
\end{equation}
For simplicity, let us restrict to the lattice $\epsilon = 0$.\footnote{In appendix \ref{app:calcs for generic epsilon} we show that in fact for any $\epsilon \in [0,4)$ we obtain the same result for $\sum_{\nu \in \mathcal{L}_\epsilon}\bar{\Psi}_{\Lambda}(\nu) \Psi_{\Lambda^{\prime}}(\nu)$.} In that case, $\nu/(4\lambda) \in \mathbb{Z}$, and the sum over $\nu$ has three kinds of terms:
\begin{align}
    S_1 &= \sum_{n = -\infty}^{\infty} \cos(\alpha_{\Lambda}\, n)\cos(\alpha_{\Lambda^\prime}\, n)\,,\\
    S_2 &= \sum_{n = -\infty}^{\infty} \cosh(\beta_{\Lambda}\, n)\cosh(\beta_{\Lambda^\prime}\, n)\,,\\
    S_3 &= \sum_{n = -\infty}^{\infty} (-1)^n \cos(\alpha_{\Lambda}\, n)\cosh(\beta_{\Lambda^\prime}\, n)\,.
\end{align}
Since $S_2$ is a sum over positive and exponentially growing terms, we see that $\Lambda>\Lambda_p$ states cannot be normalised. We hence cut off the integrals at $\Lambda=\Lambda_p$ and we do not get any contributions to $S_3$.
Focusing on $S_1$, let us consider the following representation for the Dirac delta:
\begin{equation}\label{eq:deltaexp}
    \delta(x) = \frac{1}{2\pi} \sum_{n = -\infty}^{\infty} e^{\im n x}\,,\quad x\in[-\pi,2\pi)\,.
\end{equation}
Since the value $\alpha_\Lambda=\pi$ is also possible in the limiting case $\Lambda=\Lambda_p$, there is in principle an extra contribution for $x=2\pi$, however this is an isolated point (of measure zero) which will not contribute to the norm of the state and which can hence be ignored here.

Then, expanding the product of cosines in complex exponentials leads to
\begin{equation}
    S_1 = \pi \left[ \delta(\alpha_{\Lambda}+\alpha_{\Lambda^\prime}) + \delta(\alpha_{\Lambda}-\alpha_{\Lambda^\prime}) \right].
\end{equation}
Since $\alpha_\Lambda>0$ for all $\Lambda<\Lambda_p$, the term $\delta(\alpha_{\Lambda}+\alpha_{\Lambda^ \prime})$ will always vanish, and we are left with only the $\delta(\alpha_{\Lambda}-\alpha_{\Lambda^ \prime})$ term, which we can rewrite as 
\begin{equation}
    S_1 = \pi \sqrt{\Lambda\left(\Lambda_p-\Lambda\right)}\; \delta(\Lambda-\Lambda^{\prime})\,.
\end{equation}

Then norm of a generic state \eqref{eq:norm} can thus be simplified to \begin{align}
    \left( \Psi, \Psi \right) &= 4\pi\lambda \int_0^{\Lambda_p} {\rm d}\Lambda\; |A(\Lambda)|^2 \sqrt{\Lambda\left(\Lambda_p-\Lambda\right)}\nonumber\\
    &= \frac{4 \pi}{\gamma}\int_0^{\Lambda_p} {\rm d}\Lambda\; |A(\Lambda)|^2 \sqrt{3\Lambda-\Lambda^2 \gamma^2 \lambda^2},
\end{align}
where in the second equality we have inserted the explicit form for $\Lambda_p$. Here it is easy to see that in the continuum limit ($\lambda \rightarrow 0$) we recover \eqref{eq:normWdW}, as in that case $\Lambda_p \rightarrow \infty$ and the second term in the square root vanishes; notice that the inner product now also includes the effective cosmological constant $\Lambda_{{\rm eff}}<\Lambda$ introduced below (\ref{modifiednu}).
 
Now, we are particularly interested in tracking the expectation value of $\hat{V} = \frac{\gamma\kappa\hbar}{4} |\hat{\nu}|$. Let us consider first
\begin{equation}
    \left( \Psi, \hat{|\nu|} \Psi \right) =  4\lambda \sum_{\nu \in \mathcal{L}_\epsilon}|\nu| \left|\Psi(\nu, T)\right|^2.
\end{equation}
Restricting again to the lattices with $\epsilon = 0$ for simplicity we can write this as
\begin{equation}
    \left( \Psi, \hat{|\nu|} \Psi \right) =  16\lambda^2 \sum_{n \in \mathbb{Z}}|n| \left|\Psi(4\lambda n, T)\right|^2.
\end{equation}
In this case we cannot simplify the discrete sum before computing the integrals inside $\left|\Psi(4\lambda n, T)\right|^2$, so we need to compute first the state
\begin{equation}\label{eq:psiLQC}
    \Psi(4\lambda n,T) = \int_0^{\Lambda_p} {\rm d}\Lambda\; A(\Lambda)\; e^{\im\frac{\Lambda T}{\hbar\kappa}} \cos(n\alpha_{\Lambda})
\end{equation}
(possibly numerically) for a given profile $A(\Lambda)$. 

The resulting expectation values for Gaussian states \eqref{eq:GaussianLambda} are given in figure \ref{fig:expv}, where we compare with expressions obtained in the Wheeler--DeWitt (continuum) limit (WdW) and with the classical trajectory. The qualitative features of both cases are very similar, with corrections coming from LQC only appearing at {\em large} volumes, perhaps counterintuitively. This large-volume effect can entirely be accounted for by noticing that the LQC corrections induce an effective lowering of the cosmological constant from its ``bare'' value, see the discussion below (\ref{modifiednu}). As the classical solution has (small) constant curvature, there are no high-curvature regions where we would expect holonomy corrections to be large as they are in standard LQC models when the Big Bang is replaced by a bounce. The conclusions regarding the horizon -- sudden appearance of strong quantum effects and deviation from the classical solution -- are only due to the imposition of unitarity in unimodular time, and hence identical in LQC and Wheeler--DeWitt theory. 

For a simpler comparison with the Wheeler--DeWitt case, one can change the integration variable $u = \sqrt{\Lambda}$ and perform an approximation of $\alpha_{\Lambda}$ for small $\sqrt{\Lambda}-\sqrt{\Lambda_0}$:
\begin{equation}
    \alpha_{\Lambda} = \alpha_{\Lambda_0} + 2 \sqrt{\Lambda_0} \alpha^{\prime}_{\Lambda_0} \left(u-\sqrt{\Lambda_0}\right) + \mathcal{O}\left(u-\sqrt{\Lambda_0}\right)^2,
\end{equation}
where the prime denotes derivative with respect to $\Lambda$. Then
\begin{equation}
    \int_0^{\Lambda_p} {\rm d}\Lambda\; A(\Lambda)\; e^{\im\frac{\Lambda T}{\hbar\kappa}} e^{\im n \alpha_{\Lambda}} \approx 2 N \int_0^{\sqrt{\Lambda_p}} {\rm d}u\, u\, e^{-f(T) u^2 + \tilde{g}(n) u + k}
\end{equation}
where $f(T)$ is the same as in \eqref{eq:f(T)} and
\begin{align}
    \tilde{g}(n) &\equiv \frac{u_0}{\sigma^2} + 2 {\rm i}\,n \sqrt{\Lambda_0} \alpha^{\prime}_{\Lambda_0}\,,\\
    k &\equiv -\frac{u_0^2}{2\sigma^2} + \im\, n \left(\alpha_{\Lambda_0} - 2\Lambda_0 \alpha^{\prime}_{\Lambda_0}\right)\,.
\end{align}
These expressions can now refer to the LQC case with $\alpha_\Lambda$ given in  (\ref{eq:alphabetadef}) or to the Wheeler--DeWitt case, which would correspond to $\alpha_\Lambda=\lambda\sqrt{4\gamma^2\Lambda/3}$ (compare with (\ref{eq:WdWstates int u})).

Then \eqref{eq:psiLQC} can be approximated as
\begin{align}
    \Psi(4\lambda n,T) &\approx N \frac{1}{f}\bigg\{e^{k} + e^{\bar{k}}+ \frac{1}{2}\sqrt{\frac{\pi}{f}}\bigg(\tilde{g} e^{\frac{\tilde{g}^2}{4f}}\left[\text{erf}\left(\sqrt{f \Lambda_p} - \frac{\tilde{g}}{2\sqrt{f}}\right)+\text{erf}\left(\frac{\tilde{g}}{2\sqrt{f}}\right)\right]\nonumber\\
    & + \bar{\tilde{g}}e^{\frac{\bar{\tilde{g}}^2}{4f}}\left[\text{erf}\left(\sqrt{f \Lambda_p} - \frac{\bar{\tilde{g}}}{2\sqrt{f}}\right)+\text{erf}\left(\frac{\bar{\tilde{g}}}{2\sqrt{f}}\right)\right]\bigg)\bigg\}.
\end{align}

This expression can be compared with the earlier (\ref{eq:3.13}) for the Wheeler--DeWitt case. We have found that one can use this approximation to generate approximate expectation values as given in Figure \ref{fig:expv} with results that are indistinguishable from the ones obtained using the full numerical solution.

\section{Conclusions}
\label{sec:concl}

The minisuperspace model we have studied is very simple, only including a single dynamical degree of freedom corresponding to the cosmological scale factor. We could perform a number of analytical calculations both in Wheeler--DeWitt theory and in LQC, even though the calculation of expectation values of the volume had to be done numerically. The main results are not surprising at face value: given that the classical solutions for $\nu$ are simply linear in unimodular time, for a semiclassical state the expectation value of $|\hat\nu|$ follows the classical solution all the way to the classical kink where $\nu$ changes sign. The quantum uncertainty principle does not allow for a kink in the quantum solution and the classical solution gets smeared out, and $\langle|\hat\nu|\rangle$ remains bounded away from zero. Assuming that we were only able to measure $|\hat\nu|$, in this region we would suddenly experience a strong relative deviation from the classical solution ($\langle|\hat\nu|\rangle \gg |\nu|_{{\rm cl}})$ and strong quantum fluctuations $\Delta |\nu|\sim\langle|\hat\nu|\rangle$. 

In the context of relativity, all this seems troubling since the classical horizon, the instant of time where the spatial volume goes to zero, is foliation-dependent and has no generally covariant geometric interpretation. Assuming that quantum theory and general covariance were compatible, we would require that nothing special could possibly happen here. What we are seeing in this simple example is the well-known but not always appreciated fundamental clash between the notion of unitarity and general covariance: unitarity can only apply to particular, somehow preferred foliations and time coordinates. Here we assumed unitarity in a given foliation, and one could argue that the unphysical consequences of such a demand mean that asking for unitarity is not reasonable in this context. However, this does leave us with the general question of when and how we can ask for unitarity in quantum gravity. Our unimodular time is as well-behaved as it gets, always globally defined and guaranteed to be monotonic, so it seems like it cannot be the culprit. Of course the foliation we have chosen could be considered unsatisfactory classically since it does not cover the entire spacetime, but using this criterion would seem to suggest we require knowledge of a global solution before deciding whether unitarity can be expected, something that seems unpractical outside of the simplest analytical solutions.

Finally, we saw that using LQC rather than Wheeler--DeWitt theory has only minor quantitative impact on our results, with no new conceptual insights. If unitarity in unimodular time is assumed, we still observe strong quantum effects in a region of arbitrarily low curvature. The reason why this behaviour seems to not have been observed previously is that almost the entire LQC literature  uses a massless scalar field as clock, with \cite{Chiou:2010ne,Giesel:2020raf} as notable exceptions, and solutions without matter are usually not considered. The massless scalar field can also lead to presumably clock-dependent strong quantum effects in the infinite future of de Sitter space \cite{Pawlowski:2011zf}, so again it is perhaps not surprising to see similar low-curvature quantum behaviour here. A particular form of unitarity -- assuming a self-adjoint Hamiltonian operator in a particular gauge given by a choice of lapse -- seems to be built into the constructions of LQC, and hence the clash between unitarity and general covariance also affects such models.

\

{\bf Acknowledgements}: This work was funded by the Royal Society through the University Research Fellowship Renewal URF$\backslash$R$\backslash$221005.

\appendix
\section{Norm of LQC states for generic lattices $\mathcal{L}_{\epsilon}$} \label{app:calcs for generic epsilon}

In the main text we have calculated the norm of LQC states when restricting to the $\epsilon = 0$ lattice. Here we show that the result is the same for generic $\epsilon$. Let us start by writing
\begin{equation}
    \sum_{\nu \in \mathcal{L}_{\epsilon}} f\left( \frac{\nu}{4\lambda} \right) = \sum_{n \in \mathbb{Z}} f\left( \tilde{\epsilon} + n \right), \qquad \tilde{\epsilon} \in [0,1/\lambda).
\end{equation}

This way, $S_1$ can be written as
\begin{align}
    S_1 &= \sum_{n \in \mathbb{Z}} \cos\left(\alpha_{\Lambda}n + \alpha_{\Lambda}\tilde{\epsilon}\right) \,
            \cos\left(\alpha_{\Lambda^\prime}n + \alpha_{\Lambda^\prime}\tilde{\epsilon}\right)\\
    &= \cos\left(\alpha_{\Lambda}\tilde{\epsilon}\right)\cos\left(\alpha_{\Lambda^\prime}\tilde{\epsilon}\right) \sum_{n \in \mathbb{Z}} \cos\left(\alpha_{\Lambda}n\right)\cos\left(\alpha_{\Lambda^\prime}n\right) 
    + \sin\left(\alpha_{\Lambda}\tilde{\epsilon}\right)\sin\left(\alpha_{\Lambda^\prime}\tilde{\epsilon}\right) \sum_{n \in \mathbb{Z}} \sin\left(\alpha_{\Lambda}n\right)\sin\left(\alpha_{\Lambda^\prime}n\right) \nonumber\\
    &+ \cos\left(\alpha_{\Lambda}\tilde{\epsilon}\right)\sin\left(\alpha_{\Lambda^\prime}\tilde{\epsilon}\right) \sum_{n \in \mathbb{Z}} \cos\left(\alpha_{\Lambda}n\right)\sin\left(\alpha_{\Lambda^\prime}n\right)
    + \sin\left(\alpha_{\Lambda}\tilde{\epsilon}\right)\cos\left(\alpha_{\Lambda^\prime}\tilde{\epsilon}\right) \sum_{n \in \mathbb{Z}} \sin\left(\alpha_{\Lambda}n\right)\cos\left(\alpha_{\Lambda^\prime}n\right),\nonumber
\end{align}
where in the second equality we used $\cos(x+y)=\cos(x)\cos(y)-\sin(x)\sin(y)$. Now the sum in the first term is directly $S_1|_{\epsilon=0} = \pi \delta\left(\alpha_{\Lambda}-\alpha_{\Lambda^\prime}\right)$ and it is straightforward to see that the last two terms vanish since they are sums of odd functions. Let us expand the sum in the second term and make use of the representation of the Dirac delta \eqref{eq:deltaexp} to find
\begin{align}
    \sum_{n \in \mathbb{Z}} \sin\left(\alpha_{\Lambda}n\right)\sin\left(\alpha_{\Lambda^\prime}n\right) &= \frac{1}{2}\sum_{n \in \mathbb{Z}}\left\{\cos\left[\left(\alpha_{\Lambda}-\alpha_{\Lambda^\prime}\right)n\right]-\cos\left[\left(\alpha_{\Lambda}+\alpha_{\Lambda^\prime}\right)n\right]\right\}\nonumber\\
    &= \pi \left[\delta\left(\alpha_{\Lambda}-\alpha_{\Lambda^\prime}\right)-\delta\left(\alpha_{\Lambda}+\alpha_{\Lambda^\prime}\right)\right].
\end{align}
Again, since $S_1$ is evaluated in the regime where $\Lambda,\Lambda^{\prime} < \Lambda_p$, the second delta does not contribute and we are left with
\begin{align}
    S_1 &= \pi \left[\cos\left(\alpha_{\Lambda}\tilde{\epsilon}\right)\cos\left(\alpha_{\Lambda^\prime}\tilde{\epsilon}\right) + \sin\left(\alpha_{\Lambda}\tilde{\epsilon}\right)\sin\left(\alpha_{\Lambda^\prime}\tilde{\epsilon}\right)\right]\delta\left(\alpha_{\Lambda}-\alpha_{\Lambda^\prime}\right)\nonumber\\
    &= \pi \delta\left(\alpha_{\Lambda}-\alpha_{\Lambda^\prime}\right)\;= S_1|_{\epsilon=0}.
\end{align}
Finally, as in the case of $\epsilon = 0$, the sums $S_2$ and $S_3$ are still ignored as they include real exponentials and so the functions are not normalisable for $\Lambda$ or $\Lambda^{\prime} > \Lambda_p$.

\bibliographystyle{JHEP}
\bibliography{bib}

\end{document}